\documentclass{elsart}
\usepackage{natbib}
\usepackage{graphicx}
\usepackage{psfig}

\newcommand{\arcsec}{\hbox{$^{\prime\prime}$}}

\begin{document}
\runauthor{F. Mueller S\'anchez et al.}
\begin{frontmatter}
\title{Near IR diffraction-limited integral-field SINFONI spectroscopy
of the Circinus galaxy} 
\author{F. Mueller S\'anchez, R.~I. Davies, F. Eisenhauer,}
\author{L.~J. Tacconi, and R. Genzel}

\address{Max-Planck-Institut f\"ur extraterrestrische Physik,
        Postfach 1312, D-85741 Garching, Germany}

\begin{abstract}
Using the adaptive optics assisted near infrared integral field spectrometer 
SINFONI on the VLT, we have obtained observations of the Circinus galaxy 
on parsec scales. The morphologies of the H$_2$ (1-0)\,S(1) 2.12$\mu$m 
and Br$\gamma$ 2.17$\mu$m emission lines are only slightly different, 
but their velocity maps are similar and show a gradient along 
the major axis of the galaxy, consistent with rotation.
Since $V_{\mathrm{rot}}$/$\sigma$ $\approx$ 1 suggests that
random motions are also important, it is likely that the lines arise in a
rotating spheroid or thickened disk around the AGN.
Comparing the Br$\gamma$ flux to the stellar continuum 
indicates that the star formation in this region began $\sim$10$^8$ yr ago.
We also detect the [Si{\sc vi}] $1.96\mathrm{\mu m}$,
[Al{\sc ix}] $2.04\mathrm{\mu m}$ 
and [Ca{\sc viii}] $2.32\mathrm{\mu m}$ coronal lines.
In all cases we observe a broad blue wing,
indicating the presence of two or more components in the
coronal line region. A correlation between the ionisation potential and the
asymmetry of the profiles was found for these high excitation species.
\end{abstract}

\end{frontmatter}

\section{Introduction}

The star formation history, the mass distribution, 
and the stellar and gas dynamics on scales of less than a few parsecs
in the nuclei 
of Seyfert galaxies, are some of the main debated issues in the 
context of active galactic nuclei. The Circinus galaxy, 
at a distance of $4.2\pm0.8$ Mpc (Freeman et al. \cite{freeman}),
is an ideal subject to study because of its proximity (1\arcsec\ = 20 pc). 
It is a large, highly inclined ($i=65^\circ$), spiral galaxy that hosts both 
a typical Seyfert 2 nucleus and a circumnuclear starburst. This work comprises
a summary of the results obtained after analyzing the SINFONI datacube of 
the Circinus galaxy presented in Mueller S\'anchez et al. \cite{muellersan06}

\section{The instrument: SINFONI}

\begin{figure}
    \begin{center}
     \resizebox{0.7\textwidth}{!}{\includegraphics*[clip]{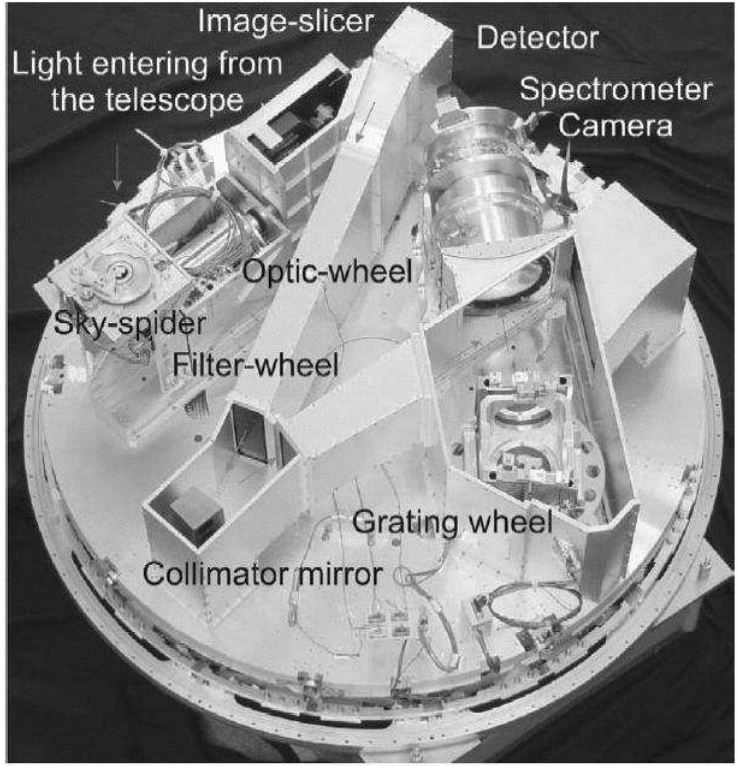}}   
    \caption{An inside view of SPIFFI (Read text for the explanation)}
    \label{spiffi}
    \end{center}
\end{figure}

SINFONI comprises an Adaptive Optics facility
(AO-Module), developed by ESO (Bonnet et al. \cite{bonnet03}) 
and SPIFFI, a NIR integral field spectrograph 
developed by MPE (Eisenhauer et al. \cite{eis03}).
The AO-Module consists of an optical relay from the telescope's
Cassegrain focus 
to the SPIFFI entrance focal plane, which includes a deformable mirror
conjugated 
to the telescope pupil. The curvature is updated on the 60 actuators
at 420 Hz, with  
a closed-loop bandwidth of 30--60 Hz,  to compensate for the
aberrations produced  
by the turbulent atmosphere. Under good atmospheric conditions an
adequate correction 
can be obtained over the full $1\times2$ arcmin$^2$ FOV with stars up
to $R = 17$ mag. 
The SPIFFI integral field spectrometer records simultaneously the
spectra of all 
image points in a two-dimensional field of view. The image scale
of SPIFFI allows sampled imaging at the diffraction limit of the telescope 
(0.0125\arcsec/pix), seeing limited observations (0.125\arcsec/pix) and
an intermediate image scale (0.05\arcsec/pix), over a FOV of
$0.8\arcsec\times0.8$\arcsec, $8\arcsec\times8$\arcsec, and
$3.2\arcsec\times3.2$\arcsec\ respectively.
The spectral resolution of the spectrometer ranges between 2000-5000 for the three
covered atmospheric bands: $J$ (1.1$\mu$m--1.4$\mu$m), $H$ (1.45$\mu$m--1.85$\mu$m)
and $K$ (1.95$\mu$m--2.45$\mu$m). The instrument is fully cryogenic, and it is
equipped with a Rockwell 2k$\times$2k HAWAII-2RG array. 
Figure~\ref{spiffi} shows an inside view of the main components of SPIFFI. 
The light enters from the top. 
The pre-optics with a filter wheel re-image the object plane from the
AO module
onto the image slicer, providing the three different image scales. The image slicer
rearranges the two dimensional field onto a one-dimensional pseudo
longslit. 
A grating wheel disperses the light and a short focal length camera then images the
spectra on the detector. After some processing of the raw data, the outcome 
is a data cube with two spatial and one spectral dimensions.

\section{Nuclear dust emission}

\begin{figure*}
    \begin{center}
     \resizebox{\hsize}{!}{\includegraphics[clip, angle=-90]{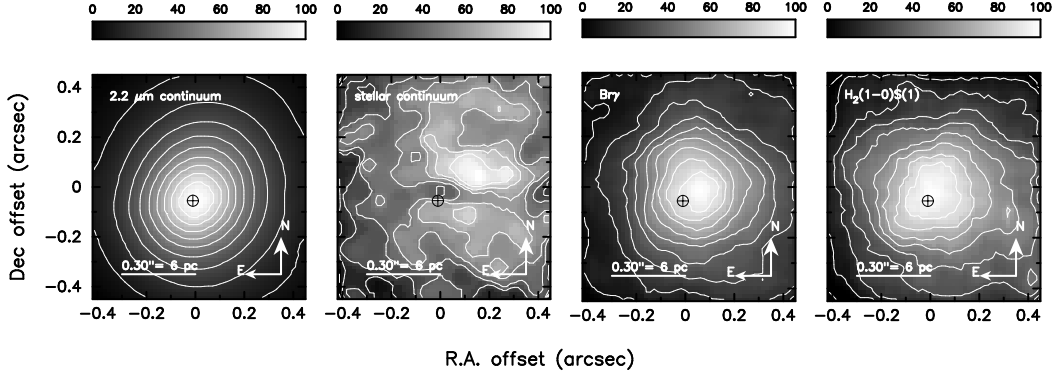}}   
    \caption{Intensity images extracted from the SINFONI data cube in the central arcsec 
             of Circinus. In each case, the gray scale extends from 0-100\% 
             of the peak flux, and contours are spaced equally between 20\% and 90\% 
             of the peak flux. An encircled cross indicates in each case the peak of 
             the continuum emission. The maps show, from left to right: %
             2.2$\mathrm{\mu m}$ continuum, stellar continuum 
             (derived from the stellar absorption bandhead $^{12}$CO(2-0)), 
             Br$\gamma$, and H$_2$(1-0)S(1).}
    \label{linemaps}
    \end{center}
\end{figure*}

Images of the 2$\mathrm{\mu m}$ continuum and the Br$\gamma$ and 
$\mathrm{H_2}$ (1-0)\,S(1) line emission, as well as the $^{12}$CO\,(2-0) 
band flux are presented in Figure~\ref{linemaps}. 
The SINFONI PSF was found to be wll represented by a symmetrical
moffat function, with a FWHM spatial resolution of 0.2\arcsec. 

In Circinus, although the AGN is highly obscured, it is revealed indirectly 
by the compact $K$-band non-stellar core. The fraction of stellar light 
in the nucleus is deduced from comparison of the $EW(^{12}$CO\,(2-0)) absorption
bandhead with star clusters models. The intrinsic equivalent width is 
pretty much independent of star formation history for an ensemble of stars as predicted 
by the models, and it has an almost constant value of $\sim12$\AA\
(Davies et al. \cite{davies06}).
The stellar continuum, which accounts for only $\sim$15\% of the total
$K$-band luminosity in the central arcsec, shows a much broader
distribution than the non-stellar part.
This extended morphology is also reflected
in the $\mathrm{H_2}$ (1-0)\,S(1) and Br$\gamma$ line profiles.   
On  these scales, less than 20 pc, an exponential profile with 
$r_{\mathrm{d}}$ = $4^{+0.5}_{-0.1}$ 
is an excellent match for each of the lines, 
consistent with the hypothesis that the stars and the molecular gas
reside in a thickened disk or rotating spheroid,  
as suggested by the kinematics (see Section~\ref{starformation}). 

We have partially resolved the non-stellar $K$-band source.
A quadrature correction of its FWHM with that of the PSF yields an
intrinsic size of $\sim$2 pc, consistent with that 
found by Prieto et al. \cite{pri04} 
and also with the sizes predicted by the unification model, in which
the hot dust  ($\sim$1000 K) lies $\sim$1 pc away from the AGN. 

\section{Star formation activity and gas kinematics} \label{starformation}

\begin{figure*}
    \begin{center}
     \resizebox{\hsize}{!}{\includegraphics[clip, angle=-90]{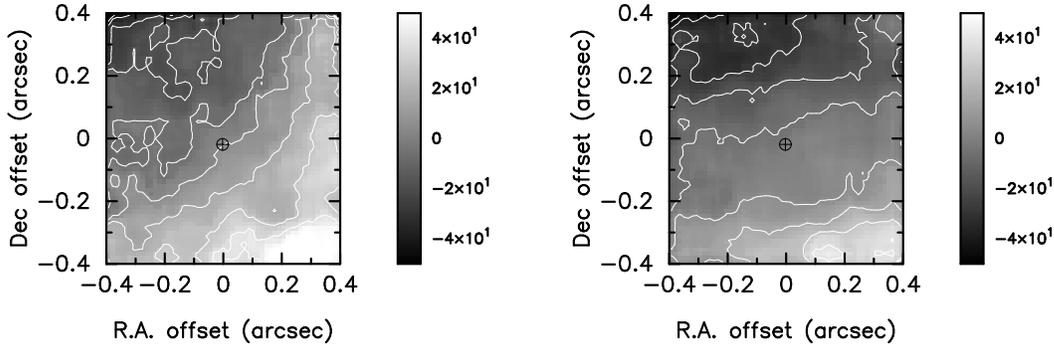}}   
    \caption{Velocity maps of the central 
             $0.8\arcsec\times0.8\arcsec$ of Circinus. 
             In each case, the gray scale extends from [-50, 50] km s$^{-1}$, 
             and contours are spaced equally every 10 km s$^{-1}$.  
             An encircled cross indicates in each case the peak of 
             the continuum emission. The maps show, \textit{Left}: H$_2$(1-0)S(1).
             \textit{Right}: Br$\gamma$}
    \label{velmaps}
    \end{center}
\end{figure*} 

As the AGN is highly obscured, no broad line region is visible.
This, together with the symmetry of the narrow line emission and its
uniform velocity field, suggests
that the Br$\gamma$ emission is associated with star formation activity 
surrounding the Seyfert nucleus. This picture is supported by the similar
morphologies in the Br$\gamma$, $^{12}$CO\,(2-0) and H$_2$ maps, 
and the consistency of the velocity fields and dispersion maps of the Br$\gamma$ 
with the H$_2$. By comparing the evolution of the $EW(\mathrm{Br\gamma})$ and
$\nu_{\mathrm{SN}}$/$L_{\mathrm{K}}^*$ with time, a moderate duration
of the star formation $t_{scl} = 10^8$ yr, and an age of $8\times 10^7$ yr 
fit best our observational constrains of $EW(\mathrm{Br\gamma})$ = 30 \AA\
and $\nu_{\mathrm{SN}}$/$L_{\mathrm{K}}^*$ = $1.47\times10^{-10}$
$L_{\odot}^{-1}$ yr$^{-1}$, indicating the presence of a young stellar population 
within a few parsecs ($R<8$ pc) of the active nucleus.

The projected velocity maps of the H$_2$ (1-0)\,S(1) and the Br$\gamma$ lines show a
velocity gradient consistent with simple rotation with the same major axis as
the large scale galaxy as can be seen in Figure ~\ref{velmaps}. 
By correcting the velocity maps for inclination ($i$ = 65$^{\circ}$), 
a rotation velocity of 75\,km\,s$^{-1}$ at 8 pc from the nucleus was found.
The intrinsic velocity dispersion of the stars in the nuclear region is almost constant 
with a value of $\sim$70\,km\,s$^{-1}$. 
Using these values we obtained an intrinsic velocity to dispersion
ratio in the maps of 
$V_{\mathrm{rot}}$/$\sigma$ $\approx$ 1.1
Because this ratio is $\sim$1, it is an indication that, in addition to
rotation, random motions are also significant in the nuclear region.

\section{The coronal lines} \label{coronal}

\begin{figure}
    \begin{center}
     \resizebox{0.48\textwidth}{!}{\includegraphics[clip]{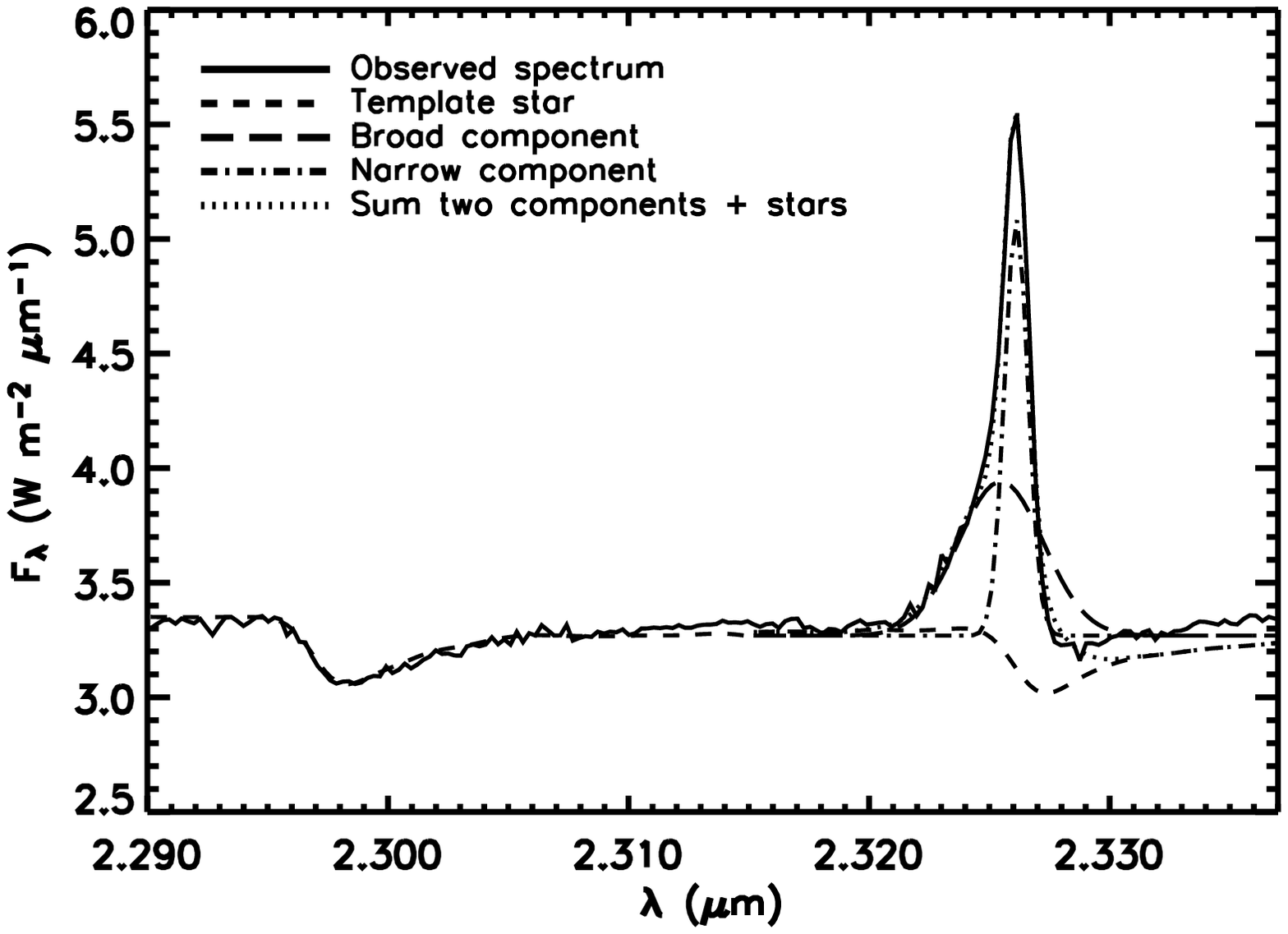}}   
     \resizebox{0.48\textwidth}{!}{\includegraphics[clip]{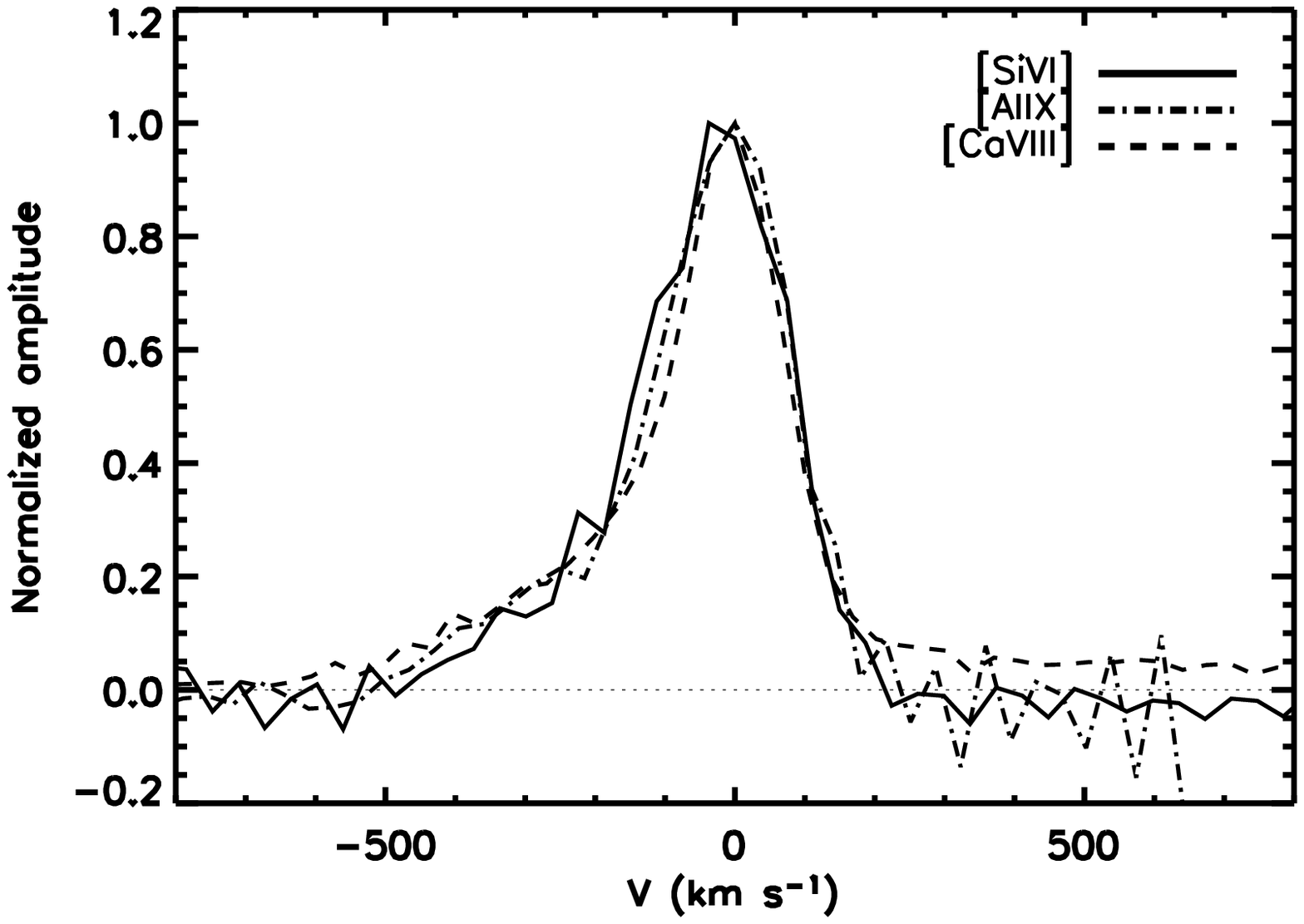}}   
    \caption{\textit{Left panel: }Nuclear spectrum around [Ca{\sc viii}] 
             showing how the stellar continuum was substracted and the 
             two components forming the line profile. \textit{Right panel: } 
             Comparison of coronal line profiles in velocity space.}
    \label{clprofiles}
    \end{center}
\end{figure}

In our observations of Circinus the highly ionised [Si{\sc vi}], [Ca{\sc viii}] and
[Al{\sc ix}] emission lines are detected. In all three profiles we observe 
asymmetric and broadened lines, 
indicating the presence of two or more components inside the coronal line region.
Indeed, the three asymmetric coronal lines were best fitted by the superposition 
of two Gaussians as can be seen in Figure~\ref{clprofiles} for the case of the 
[Ca{\sc viii}] emission line. The kinematics of the coronal gas were also analyzed 
by comparing the emission line profiles of the three coronal lines. 
Figure~\ref{clprofiles} also shows the results for the three high excitation lines 
detected with SINFONI. 
The blue shifted ($-100$ to $-200$\,km\,s$^{-1}$) 
broad ($>$300\,km\,s$^{-1}$) wing is spatially extended. 
On the other hand, the narrow component is at systemic velocity and spatially compact. 
These characteristics suggest that 
the two components originate in different regions and could even be
excited by different mechanisms.
In the case of the broad component, its blue shift indicates that part
of the gas must arise in outflows around the AGN 
(Rodr\'iguez-Ardilla et al. \cite{rod04}).
The most likely scenario is that this comprises a multitude of
outflowing cloudlets producing many narrow components 
at different velocities which combine to produce the observed
morphology and profile. 
It has been suggested that similar out-flowing cloudlets in NGC\,1068
might arise from ablation of the larger clouds (Cecil et al. \cite{cecil02}).
In the case of the narrow component, the fact that it is compact, centered on 
the nucleus, and at the systemic velocity suggest that it originates
physically close to the AGN.
The narrow line width ($\sim$180\,km\,s$^{-1}$) indicates that they
must be excited by photoionization from a hard ionising continuum
(Oliva \cite{oliva99}).

Interestingly, we observe correlations between the ionisation
potential (IP) and both the blueshift and relative strength of the
broad wing, indicating that different species (with different IP)
originate in different regions.
We suggest that the higher ionization species have faster outflow
velocities because they are closer to the AGN where the radiative
acceleration is stronger 
(and the cloudlets have not been slowed by
travelling a long distance through the
interstellar medium), and weaker fluxes because there are fewer
photons hard enough to ionise them.

\end{document}